\documentclass{acm_proc_article-sp}
\usepackage{url}
\usepackage{fixltx2e}
\usepackage{rotating}
\usepackage{color}
\usepackage{array}
\definecolor{grey}{rgb}{0.95, 0.95, 0.95}
\usepackage[final]{listings}
\usepackage[font=footnotesize]{subfig}
\usepackage{sidecap}

\newcommand{\GUPS}{\mbox{GUP/s}}
\newcommand{\MBS}{\mbox{MB/s}}
\newcommand{\GBS}{\mbox{GB/s}}

\newcommand{\FLOP}{\mbox{flop}}
\newcommand{\BYTE}{\mbox{byte}}
\newcommand{\BYTES}{\mbox{bytes}}
\newcommand{\GFS}{\mbox{Gflop/s}}
\newcommand{\DB}{\mbox{dB}}
\newcommand{\muop}{\mbox{$\mu$OP}}
\newcommand{\bq}{\begin{equation}}
\newcommand{\eq}{\end{equation}}
\renewcommand{\arraystretch}{1.2}
\graphicspath{{figures/}}
\DeclareCaptionType{copyrightbox}
\def\rabbitct{R{\sc abbit}CT}

\begin{document}

\title{Pushing the limits for medical image reconstruction on recent standard multicore processors}

\numberofauthors{5} 
\author{
\alignauthor
Jan Treibig\\
       \affaddr{Erlangen Regional Computing Center}\\
       \affaddr{Martensstr. 1}\\
       \affaddr{91058 Erlangen, Germany}\\
       \email{jan.treibig@rrze.uni-erlangen.de}
\alignauthor
Georg Hager\\
       \affaddr{Erlangen Regional Computing Center}\\
       \affaddr{Martensstr. 1}\\
       \affaddr{91058 Erlangen, Germany}\\
       \email{georg.hager@rrze.uni-erlangen.de}
\alignauthor 
Hannes G. Hofmann\\
       \affaddr{Pattern Recognition Lab}\\
       \affaddr{Martensstr. 3}\\
       \affaddr{91058 Erlangen, Germany}\\
       \email{hannes.hofmann@informatik.uni-erlangen.de}
\and
\alignauthor 
Joachim Hornegger\\
       \affaddr{Pattern Recognition Lab}\\
       \affaddr{Martensstr. 3}\\
       \affaddr{91058 Erlangen, Germany}\\
       \email{jh@informatik.uni-erlangen.de}
\and
\alignauthor 
Gerhard Wellein\\
       \affaddr{Erlangen Regional Computing Center}\\
       \affaddr{Martensstr. 1}\\
       \affaddr{91058 Erlangen, Germany}\\
       \email{gerhard.wellein@rrze.uni-erlangen.de}
}
\date{30 July 1999}

\maketitle
\begin{abstract}
  Volume reconstruction by backprojection is the computational
  bottleneck in many interventional clinical computed tomography (CT)
  applications. Today vendors in this field replace special purpose
  hardware accelerators by standard hardware like multicore chips and
  GPGPUs.  Medical imaging algorithms are
  on the verge of employing High Performance Computing (HPC) technology, and 
  are therefore an interesting new candidate for optimization.
  This paper presents low-level optimizations for the
  backprojection algorithm, guided by a thorough performance analysis
  on four generations of Intel multicore processors (Harpertown,
  Westmere, Westmere EX, and Sandy Bridge).

  We choose the \rabbitct{} benchmark, a standardized testcase well
  supported in industry, to ensure transparent and comparable results.
  Our aim is to provide not only the fastest possible implementation but also
  compare to performance models and hardware counter data in order to
  fully understand the results. We separate the influence of
  algorithmic optimizations, parallelization, SIMD vectorization, and
  microarchitectural issues and pinpoint problems with
  current SIMD instruction set extensions on standard CPUs (SSE, AVX).
  The use of assembly language is mandatory for best performance. 
  Finally we compare our results to the best GPGPU implementations
  available for this open competition benchmark.
\end{abstract}

\section{Introduction and Related Work}
\label{sec:intro}

\subsection{Computed tomography}

Computed tomography (CT) \cite{kak01-POC} is nowadays an established technology to
non-invasively determine a three-dimensional (3D) structure from a series of 2D
projections of an object. Beyond its
classic application area of static analysis in clinical environments
the use of CT has accelerated  substantially in recent years,
e.g., towards material science or time-resolved scans supporting
interventional cardiology. The numerical volume reconstruction scheme
is a key component of modern CT systems and is known to be very
compute-intensive. Acceleration through special-purpose hardware such
as FPGAs is a typical approach 
to meet the constraints of real-time processing. Integrating
nonstandard hardware into commercial CT systems adds considerable
costs both in terms of hardware and software development, as well as system
complexity. From an economic view the use of standard x86 processors
would thus be preferable. Driven by Moore's law the compute
capabilities of standard CPUs have now the potential to meet the
requested CT time constraints.
\begin{figure}[htb]
\includegraphics[width=\linewidth]{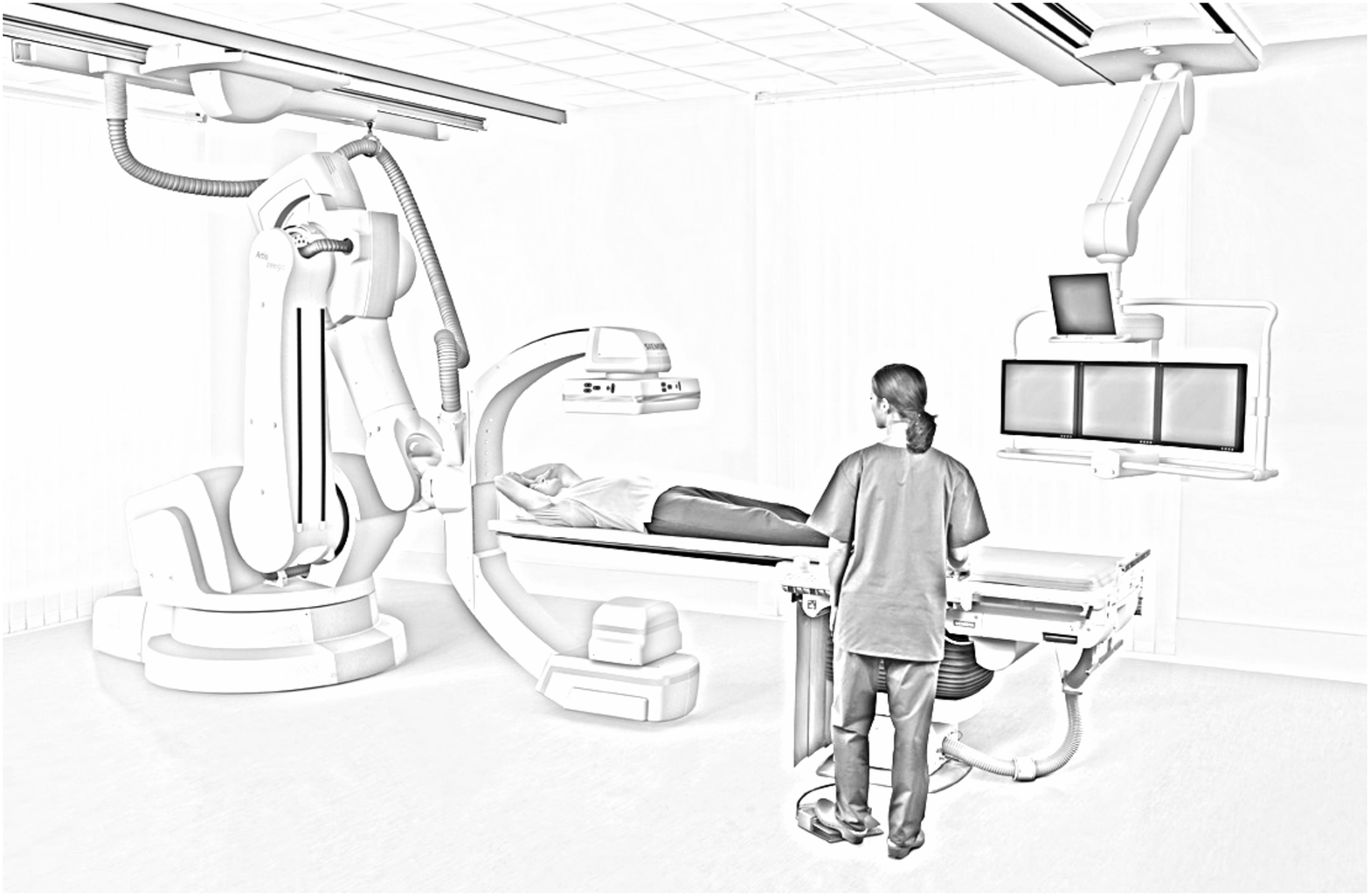}
\caption{C-arm system illustration (Axiom Artis Zeego, Siemens Healthcare, Forchheim, Germany).}
\label{fig:carm}
\end{figure}

The volume reconstruction step for recent  C-arm systems
with flat panel detector can be considered as a prototype for modern
clinical CT systems. Interventional C-arm CTs, such as the one
sketched in Fig.~\ref{fig:carm}, perform the rotational acquisition of
496 high resolution X-ray projection images (1248$\times$960 pixels) in 20
seconds~\cite{Strobel09-3IW}. This acquisition phase sets a constraint
for the maximum reconstruction time to attain real-time
reconstruction. In practice filtered backprojection (FBP) methods such
as the Feldkamp algorithm~\cite{Feldkamp84-PCA} are widely used for
performance reasons. The algorithm consists of 2D pre-processing
steps, backprojection, and 3D post-processing. Volume data is
reconstructed in the backprojection step, making it by far the most
time consuming part~\cite{Heigl07-HSR}. It is characterized by high
computational intensity, nontrivial data dependencies, and complex
numerical evaluations but also offers an inherent embarrassingly
parallel structure. In recent years hardware-specific optimization of
the Feldkamp algorithm has focused on
GPUs~\cite{Mueller98-R3C,Mueller07-WDC} and IBM Cell
processors~\cite{Scherl07-IOT}. For GPUs in particular, large
performance gains compared to CPUs were
reported~\cite{Mueller07-WDC} or documented by the standardized
\rabbitct{} benchmark~\cite{Rohkohl09-RAO,hpc:RabbitCT:2011}.
Available studies with standard CPUs indicate that large servers are
required to meet GPU performance~\cite{Hofmann11-CPO}. 
In this report we also use the \rabbitct{} environment, 
which defines a clinically relevant test
case and is supported by industry.
\begin{figure}[tb]
        \subfloat[Skin]{%
        \includegraphics[width=0.49\linewidth]{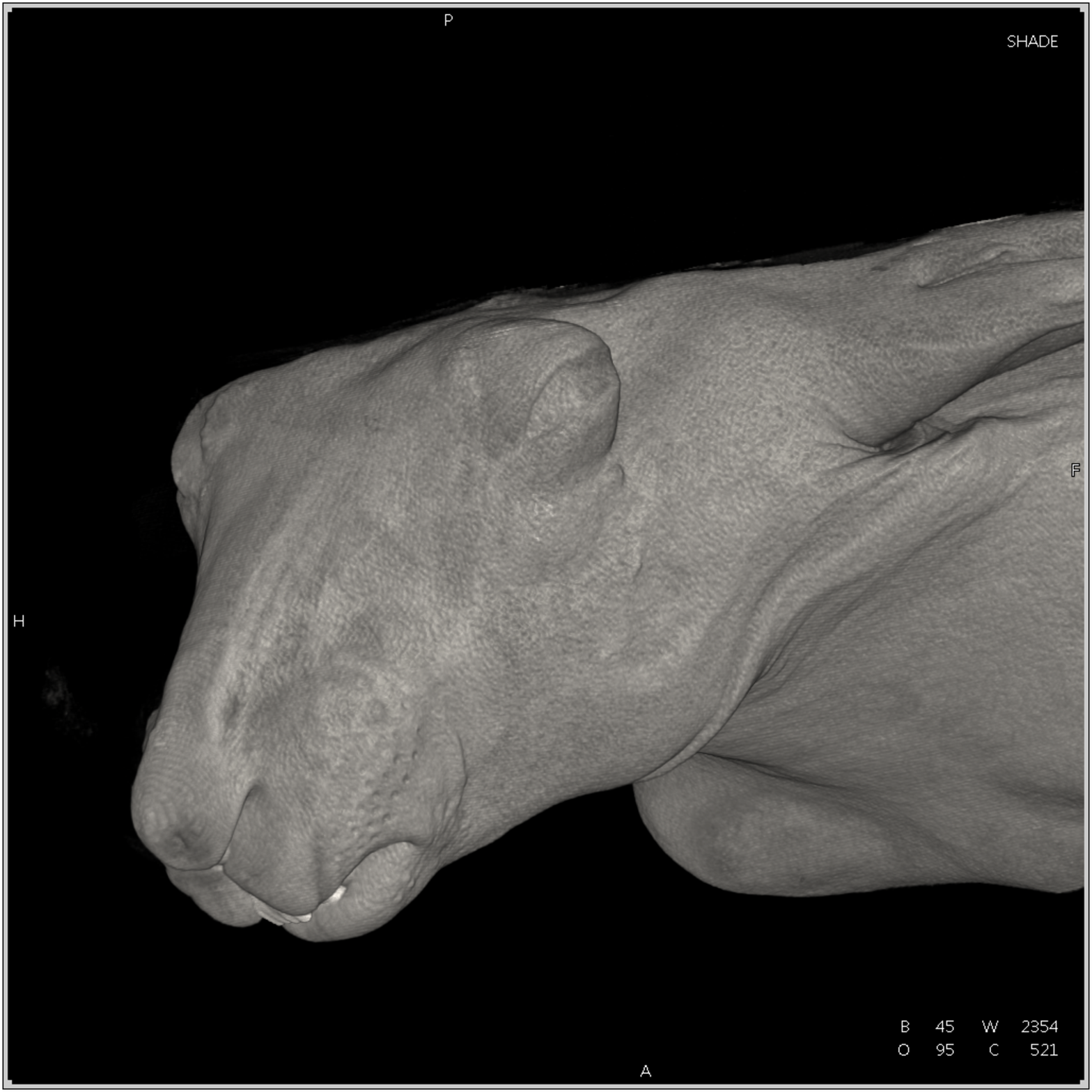}
        \label{fig:rabbit_skin}}
        \subfloat[Bones]{%
        \includegraphics[width=0.49\linewidth]{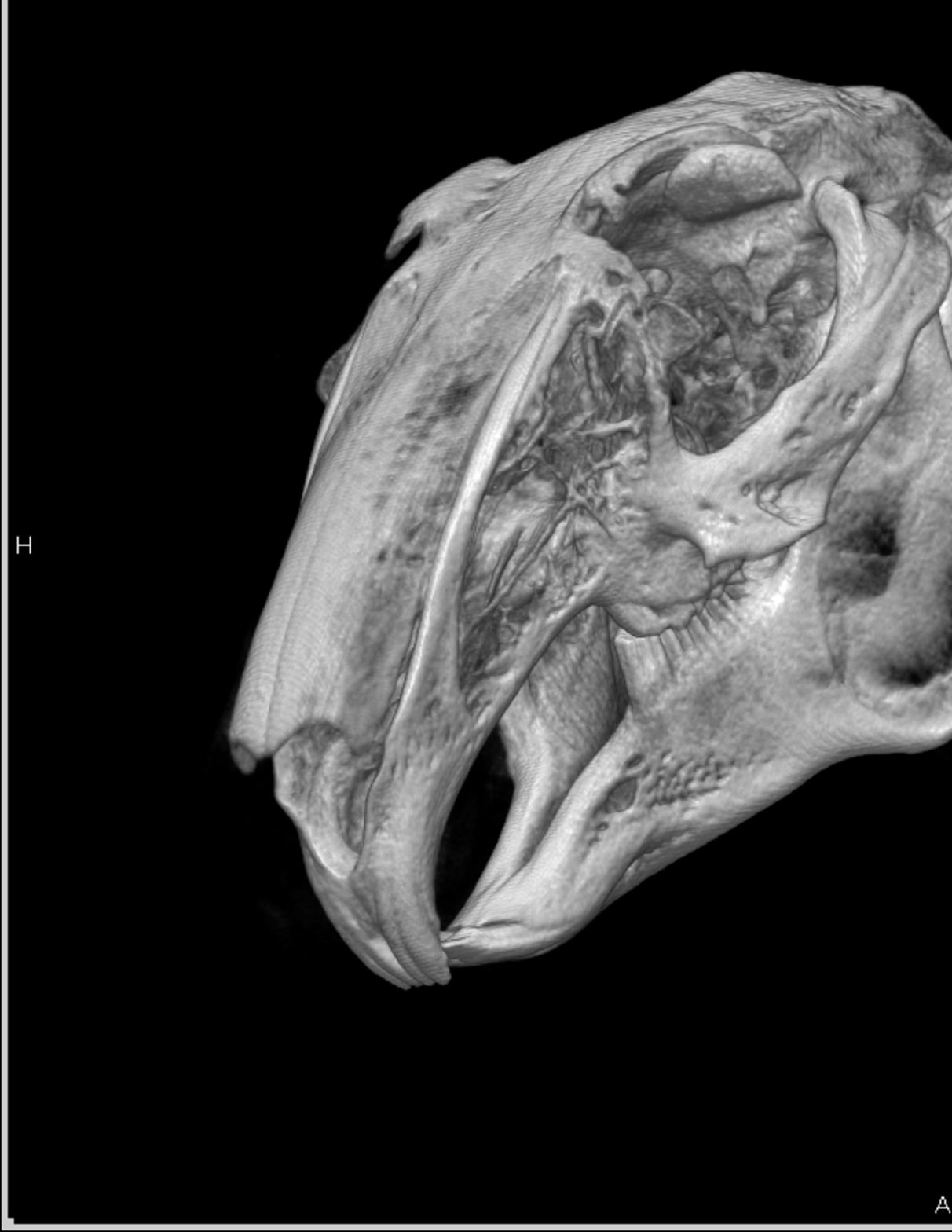}%
        \label{fig:rabbit_bones}}\caption{\label{fig:rabbit}Volume
        renderings based on the reconstruction of 2D X-ray projections 
        of a rabbit.}
\end{figure}
\rabbitct{} is an open competition benchmark based on C-arm CT images of a rabbit
(see Fig.~\ref{fig:rabbit}). It allows to compare the manifold of
existing hardware technologies and implementation alternatives for
reconstruction scenarios by applying them to a fixed, well-defined problem.

At research level, recent reconstruction methods use more advanced iterative
techniques, which can provide superior image quality in special cases like
sparse or irregular data~\cite{Kunze07-IEF}. Other algorithms are used to
reconstruct time-resolved volumes (``3D+$t$'')~\cite{Rohkohl09-I4M}, e.g., for
cardiac imaging. However, both approaches incorporate several backprojection
steps, making performance improvements on the \rabbitct{} benchmark valuable
for them as well. The same holds for industrial CTs in material science used in
nondestructive testing (NDT), which additionally run at higher resolution and thus further
increase the computational requirements~\cite{Keck09-HRI}. The high computational demand of
backprojection algorithms together with the growing acceptance of parallel
computing in medical applications make them interesting new candidates 
on the interface to High Performance Computing.

\subsection{Modern processors}

The steadily growing transistor budget is still pushing forward the
compute capabilities of commodity processors at rather constant
price and clock speed. Improvement and scaling of established
architectural features in the core  (like, e.g., SIMD and
SMT; see below for details) in addition to
increasing the number of cores per chip lead to peak performance
levels which enable standard CPUs  to meet the time
constraints of interventional CT systems. Optimized single core
implementations are thus mandatory for reconstruction
algorithms. Complemented by standard optimization techniques, a
highly efficient SIMD (Single Instruction Multiple Data) code is the
key performance component. Owing to the nontrivial data parallelism
in the reconstruction scheme, SIMD optimizations need to be done
at low level, i.e., down to assembly language. However, these efforts will
pay off for future computer architectures since an efficient SIMD
implementation will become of major importance to further benefit from
increasing peak performance numbers.

Scaling SIMD width is a safe bet to increase raw core performance
and optimize energy efficiency (i.e., maximize the \FLOP/Watt ratio),
which is known to be the critical component in future HPC systems.
Recently  Intel has played the game by doubling the SIMD width
from the SSE instruction set (128-bit registers) to 
AVX~\cite{avxManual} (256-bit registers, with larger widths
planned for future designs),
which is implemented in the ``Sandy Bridge'' architecture. 
More ongoing projects are pointing in the same direction,
like Intel's Many Integrated Core (MIC)~\cite{hpc:IntelMIC:2011}
architecture or the Chinese Godson-3B chip~\cite{hpc:Godson3:2011}.
Wider SIMD units do not change core complexity substantially
since the optimal instruction throughput does not depend on the SIMD
width. However, the benefit in the application strongly depends on the
level and the structure of data parallelism as well as the capability
of the compiler to detect and exploit it. Compilers are
very effective for simple code patterns
like streaming kernels~\cite{hpc:StreamBench:2011}, 
while more complex loop structures require manual
intervention, at least to the level of compiler intrinsics.
This is the only safe way to utilize SIMD capabilities to their full potential. 
The impact of wider SIMD units on
programming style  is still unclear since this trend currently
starts to accelerate. Of course wide SIMD execution is most
efficient for in-cache codes because of the
large ``DRAM gap.''

Simultaneous Multi-Threading (SMT) is another technology to improve
core utilization at low architectural impact and energy costs. It is
obvious that SMT should be most beneficial for those ``in cache
codes'' that have limited single thread efficiency due to bubbles in
arithmetic/logic pipelines, and where cache bandwidth is not the
dominating factor. Naively one would not expect improvements from SMT
if a code is fully SIMD-vectorized since SIMD is
typically applied to simple data-parallel structures, which are a
paradigm for efficient pipeline use. Since many programmers do not
care about pipeline occupancy in their application the
benefit of SMT is often tested in an experimental way, without 
arriving at a clear explanation for why it does or does not
improve performance.

This paper is organized as follows. In Sect.~\ref{sec:code_analysis} 
we perform a first  analysis of the backprojection 
algorithm implemented in the
\rabbitct{} framework using simple
metrics like arithmetic throughput and memory bandwidth as guidelines
for estimating performance. We address processors from four generations of
Intel's x86 family (Harpertown, Westmere, Westmere EX, and Sandy
Bridge). Basic optimization rules such as minimizing 
overall work are applied. In Sect.~\ref{sec:core_optimization}
we show how to efficiently vectorize the inner loop kernel
using SSE and AVX instructions, and discuss the possible
benefit of multithreading with SMT. Sect.~\ref{sec:model}
provides an in-depth performance analysis, which will show
that simple bandwidth or arithmetic throughput models
are inadequate to estimate the performance of the algorithm.
OpenMP parallelization and related optimizations like
ccNUMA placement and bandwidth reductions are discussed
in Sect.~\ref{sec:omp}. Performance results for cores,
sockets, and nodes on all four platforms are given in 
Sect.~\ref{sec:results}, where we also interpret the effect
of the different optimizations discussed earlier and validate
our performance model. Finally we compare our results
with current GPGPU implementations in Sect.~\ref{sec:gpgpu}.

\section{Experimental testbed}
\label{sec:machines}

A selection of modern Intel x86-based multicore processors (see
Table~\ref{tab:arch}) has been chosen to test the performance potential of our
optimizations. All of these chips feature a large outer level cache, which is
shared by two (Core 2 Quad ``Harpertown''), four (Sandy Bridge), six (Westmere EP),
or ten cores (Westmere EX). We refer to the maximum number of cores sharing an outer
level L2/L3 cache as an ``L2/L3 group.'' 

With the initiation of the Core i7 architecture the memory subsystem
of Intel processors was redesigned to allow for a substantial increase in memory
bandwidth, at the price of introducing ccNUMA on
multisocket servers.
At the same time Intel also relaunched
simultaneous multithreading (SMT, a.k.a. ``Hyper-Threading'') with two
SMT threads per physical core. The most recent processor, Sandy Bridge,
is equipped with a new instruction scheduler, supports the new AVX
SIMD instruction set extension, and has a new last level cache subsystem
(which was already present in Nehalem EX). The 10-core
Intel Westmere EX  is not mainly targeted at HPC clusters but at large
mission-critical servers. It reflects the performance maximum for x86 shared-memory
nodes.  A comprehensive summary of the most important processor features is
presented in Table~\ref{tab:arch}.  Note that the Sandy Bridge model used here 
is a desktop variant, while the other processors are of the server (``Xeon'') type.
Table~\ref{tab:arch} also contains bandwidth measurements for a simple
update benchmark:
\begin{lstlisting}
  for(int i=0; i<N; ++i)
    a[i] = s * a[i];
\end{lstlisting}
This benchmark reflects the data streaming properties of the reconstruction
algorithm and is thus better suited than STREAM~\cite{hpc:StreamBench:2011} 
as a baseline for a  quantitative performance model.

We use the Intel C/C++ compiler in version 12.0; since most of our performance-critical
code is written in assembly language, this choice is marginal, however. 
Thread affinity, hardware performance
monitoring, and low-level benchmarking was implemented via the LIKWID tool 
suite~\cite{psti,likwid}, using the tools likwid-pin, likwid-perfctr, and
likwid-bench, respectively.
\begin{table*}[tbp]
    \caption{Test machine specifications. The cacheline size is 64 \BYTES\ 
	for all processors and cache levels. The update benchmark results were obtained with the likwid-bench tool.}
    \label{tab:arch}
    \centering\footnotesize
    \renewcommand{\arraystretch}{1.5}
	\begin{tabular}{lccccc}
	    \hline
	    Microarchitecture           &Intel Harpertown  &Intel Westmere   &Intel Westmere EX  &Intel Sandy Bridge \\
	    Model                       &Xeon X5482        &Xeon X5670       &Xeon E7- 4870      &Core i7-2600\\
	    Label                       &HPT               &WEM              &WEX                &SNB \\
	    \hline                                                                                                   
	    Clock [GHz]                 &3.2               &2.66 (2.93 turbo)&2.40               &3.4 (3.5 turbo)   \\
        Node sockets/cores/threads  &2/8/-             &2/12/24          &4/40/80            &1/4/8             \\
	    \hline                                                                                                  
	    Socket L1/L2/L3 cache       &4$\times$32k/2$\times$6M/- &6$\times$32k/6$\times$256k/12M     &8$\times$32k/8$\times$256k/30M       &4$\times$32k/4$\times$256k/8M       \\
	    \hline                                                                                                  
        Bandwidths [GB/s]:          &                  &                 &                   &       \\\hline 
        Theoretical socket BW       &12.8              &32.0             &34.2               &21.3               \\
	    Update (1 thread)           &5.9               &15.2             &8.3                &16.5               \\
        Update (socket)             &6.2               &20.3             &24.6               &17.3               \\
	    Update (node)               &8.4               &39.1             &98.7               &-                  \\
	    \hline
	\end{tabular}
\end{table*}
%
%
%

\section{The algorithm}
\label{sec:code_analysis}

\subsection{Code analysis}

\begin{figure}[tbp]
\includegraphics[width=\linewidth]{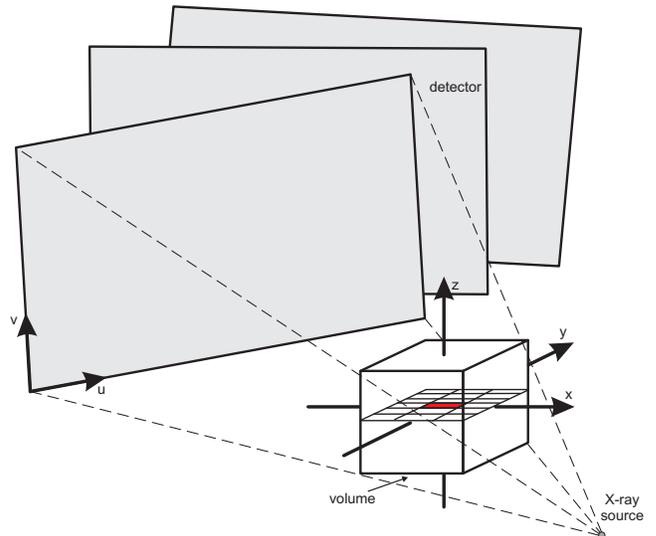}
\caption{Setup geometry for generating the CT projection images. The size of the volume 
  is always $256^3$\,mm$^3$, but the number of voxels may vary.}
\label{fig:geometry}
\end{figure}
\begin{lstlisting}[caption={Voxel update loop nest for the plain backprojection algorithm. This gets executed for each projection \texttt{I}. All variables are of type \texttt{float} unless indicated otherwise. The division into parts (see text) is only approximate since there is no 1:1 correspondence to the SIMD-vectorized code.},label=lst:alg,float=p,numbers=left,numberstyle=\tiny]
wz = offset_z;
for(int z=0; z<L; z++, wz+=MM) {
 wy = offset_y;

 for (int y=0; y<L; y++, wy+=MM) {
   wx = offset_x;
   valtl=0.0f; valtr=0.0f;
   valbl=0.0f; valbr=0.0f;

 // Part 1 --------------------------
   for (int x=0; x<L; x++, wx+=MM) {
     uw = (A[0]*wx+A[3]*wy+A[6]*wz+A[9]);
     vw = (A[1]*wx+A[4]*wy+A[7]*wz+A[10]);
     w  = (A[2]*wx+A[5]*wy+A[8]*wz+A[11]);

     u = uw * 1.0f/w; v = vw * 1.0f/w;

     int iu = (int)u, iv = (int)v;%\label{l:intc}%

     scalx = u - (float) iu;
     scaly = v - (float) iv;
 // Part 2 ---------------------------
     if (iv>=0 && iv<ISY) {%\label{l:if1}%
       if (iu>=0 && iu<ISX) 
         valtl = I[iv*ISX + iu];%\label{l:idx1}%
       if (iu>=-1 && iu<ISX-1) 
         valtr = I[iv*ISX + iu+1];%\label{l:idx2}%
     }

     if (iv>=-1 && iv<ISY-1) {%\label{l:if2}%
       if (iu>=0 && iu<ISX) 
         valbl = I[(iv+1)*ISX + iu];%\label{l:idx3}%
       if (iu>=-1 && iu<ISX-1) 
         valbr = I[(iv+1)*ISX + iu+1]; %\label{l:idx4}%
     }
 // Part 3 ---------------------------
     vall = scaly*valbl + (1.0f-scaly)*valtl;
     valr = scaly*valbr + (1.0f-scaly)*valtr;
     fx   = scalx*valr  + (1.0f-scalx)*vall;

     %VOL[z*L*L + y*L + x] += 1.0f/(w*w)*fx;\label{l:vupd}% 
   } // x
 } // y
} // z
\end{lstlisting}
The backprojection algorithm 
(as provided by the RabbitCT framework~\cite{hpc:RabbitCT:2011}, see Listing~\ref{lst:alg}) 
is usually implemented in single precision (SP) and exhibits a
streaming access pattern for most of its data traffic. 
One volume reconstruction uses 496 CT
images (denoted by \verb+I+) of 1248$\times$960 pixels each (\verb.ISX.$\times$\verb.ISY.).
The
volume size is $256^3$\,mm$^3$. \verb+MM+ is the voxel size and changes depending
on the number of voxels. The most common resolution in present clinical
applications is 512 voxels in each direction (denoted by the problem size \verb.L.).
Each CT image is accompanied by a $3\times{}4$ projection
matrix \verb.A., which projects an arbitrary point in 3D space onto the CT image.
The algorithm computes the contributions to each voxel across all projection
images. 
The reconstructed volume
is stored in array \verb+VOL+. Voxel coordinates (indices) are denoted by 
\verb+x+, \verb+y+, and \verb+z+, while pixel coordinates are
called \verb+u+ and \verb+v+. See Fig.~\ref{fig:geometry}
for the geometric setup.

The aggregated size of all projection images is $\approx${}$2.4$\,GB. The total data
transfer volume of one voxel sweep comprises the loads from the projection image
and an update operation (\verb.VOL[i]+=s., see line~\ref{l:vupd} in Listing~\ref{lst:alg}) 
to the voxel array. 
The latter incurs 8~\BYTES\ of traffic per voxel
and results (for problem size $512^3$) in a data volume of $1$\,GB, or $496$\,GB for all
projections. The traffic caused by the projection images is not
easy to quantify since it is not a simple stream; it is defined by a ``beam'' of locations
slowly moving over the projection pixels as the voxel update loop nest progresses. 
It exhibits some temporal locality since
neighboring voxels are projected on proximate pixels of the image, but there may also
be multiple streams with large strides.
On the computational side, the basic version of this algorithm performs 13 
additions, 5 subtractions, 17 multiplications, and 3 divides.

\subsection{Simple performance models}\label{sec:perfmod}

Based on this knowledge about data transfers and arithmetic operations we can
derive a first rough estimate of expected upper performance bounds. 
The arithmetic limitation results in 21 cycles per
vectorized update (4 and 8 inner loop iterations for SSE and AVX,
respectively), assuming full vectorization and a throughput of 
one divide per cycle. This takes into account that all architectures under
consideration can execute one addition and one multiplication per cycle, 
neglects the slight imbalance of additions versus multiplications, 
and assumes that the pipelined \verb.rcpps. instruction can be employed
for the divisions (see Sect.~\ref{sec:simd} for details).

On the other hand, runtime can also be estimated based on the data transfer volume
and the maximum data transfer capabilities of the nodes measured with the synthetic
update benchmark described in Sect.~\ref{sec:machines}. 
The following table shows upper performance bounds for a full $512^3$ reconstruction
based on  arithmetic and bandwidth limitations on the four systems in the testbed
(full nodes):
\begin{center}
\begin{tabular}{lcccc}
                             &HPT       &WEM     &WEX       &SNB\\\hline
        Arithm. lim. [\GUPS]&4.86      &6.75    &13.9       &5.31\\
        BW lim. [\GUPS]     &1.06      &4.90    &11.1       &2.15    \\
\end{tabular}
\end{center}
Performance is given in billions of voxel updates
per second (\GUPS),\footnote{We use SI prefixes,
i.e., 1\,\GUPS\ means $10^9$ updates per second. This is inconsistent with
a large part of the literature on medical image reconstruction, where
``G'' is used as a binary prefix for $2^{30}\approx1.074\cdot 10^9$~\cite{Goddard07-EOC}} 
where one ``update'' represents the 
reconstruction step of one voxel using a single image. 
The low arithmetic limitation for the single socket Sandy Bridge is caused by
its wide AVX vector size and its faster clock.  While above predictions
seem to indicate a strongly memory-bound situation, they are far from
accurate: The runtime is governed by the
number of instructions and the cycles it takes to execute these instructions. 
A reduction to the purely ``useful'' work, i.e., to arithmetic operations, can
not be made since this algorithm is nontrivial to vectorize due to the scattered
load of the projection image; it therefore involves many more non-arithmetic
instructions (see Sect.~\ref{sec:simd} for details). We will show later
that a more careful analysis leads to a completely different picture, and that
further optimizations can change the bottleneck analysis considerably.


In order to have a better view on low-level optimizations we divide the algorithm 
into three parts:
\begin{enumerate}
    \item Geometry computation: Calculate the index of the projection of a voxel
      in pixel coordinates
    \item Load four corner pixel values from the projection image
    \item Interpolate linearly for the update of the voxel data
\end{enumerate}

\subsection{Algorithmic optimizations}
\label{sec:algo_optimization}

The first optimizations for a given algorithm must be on a
hardware-independent level.  Beyond elementary measures like moving
invariant computations out of the inner loop body and reducing the
divides to one reciprocal (thereby reducing the \FLOP\ count to 31), 
a main optimization is to minimize the
workload. Voxels located at the corners and edges of the volume are
not visible on every projection, and can thus be ``clipped off'' and
skipped in the inner loop. This is not a new idea, but the approach
presented here improves the work reduction from
24\%~\cite{Hofmann10-ACR} to nearly 39\%.

The basic building block for all further steps is the update of a consecutive
line of voxels in $x$ direction, covered by the inner loop level in 
Listing~\ref{lst:alg}. We refer to this as the ``line update kernel.'' 
The geometry, i.e., the position of the first and the
last relevant voxel for each projection image and line of voxels
is precomputed. This information is specific for a given geometric 
setup, so it can be stored and used
later during the backprojection loop. Reading the data from memory
incurs an additional transfer volume of
$512^2${}$\times${}$496${}$\times${}$4$~\BYTES$=${}$496$\,\MBS\ (assuming
16-bit indexing), which is
negligible compared to the other traffic. The advantage of line-wise
clipping is that the shape of the clipped voxel volume is much more
accurately tracked than with the blocking approach described
in~\cite{Hofmann10-ACR}.

The conditionals (lines~\ref{l:if1} and \ref{l:if2} in Listing~\ref{lst:alg}), 
which ensure correct access to the projection image, involve no
measurable overhead for the scalar case due to the hardware branch prediction.
However, for vectorized code they are potentially costly since an appropriate
mask must be constructed whenever there is the possibility that a SIMD vector
instruction accesses data outside the projection~\cite{Hofmann10-ACR}.
To remove this complication, separate  buffers are used to hold suitably
zero-padded copies of the projection images, so that there is no need for
vector masks. The additional overhead is
far outweighed by the performance advantage for vectorized code execution.
The conditionals are also effectively removed by the clipping optimization 
described above, but we need a code version without clipping for validating
our performance model later.

Note that a similar effect could be achieved by peeling off scalar 
loop iterations to make the length of the inner loop body a multiple of the 
SIMD vector size and ensure aligned memory access. However, this may
introduce a significant scalar component especially for small problem
sizes and large vector lengths.

\section{Single core optimizations}
\label{sec:core_optimization}

For all further optimizations we choose an implementation of the line
update kernel in C as the baseline. All algorithmic optimizations from
Sect.~\ref{sec:algo_optimization} have already been applied.


The performance of present processors on the core level relies
on instruction-level parallelism (ILP) by pipelined and superscalar
execution, and data-parallel operations (SIMD). 
We also regard  simultaneous multithreading (SMT) as a single core
optimization since it is a hardware feature to increase the efficiency
of the execution units by filling pipeline bubbles with useful work:
The idea is to duplicate parts of the hardware resources (control logic,
registers) in order
to allow a quasi-simultaneous execution of different threads while sharing
other parts like, e.g., floating-point pipelines. In a sense this is an alternative
to outer loop unrolling, which can also provide multiple
independent instruction streams. 

We  elaborate on the SIMD vectorization here and comment on the 
benefit of SMT in Sect.~\ref{sec:ilpsmt}.

\subsection{SIMD vectorization}\label{sec:simd}

No current compiler is able to efficiently vectorize the backprojection 
algorithm, so we have implemented the code directly in x86 assembly
language. Using SIMD intrinsics could ease the
vectorization but adds some uncertainties with regard to register 
scheduling and hence does not allow full control over the
instruction code. 
All data is aligned to enable packed and aligned loads/stores of 
vector registers (16 (SSE) or 32 (AVX) \BYTES\ with one instruction).

The line update kernel operates on consecutive voxels.  Part 1
of the algorithm (see Sect.~\ref{sec:perfmod}) 
is straightforward to vectorize, since it is arithmetically limited
and fully benefits from the increased register width. The division is replaced by
a reciprocal. SSE provides the fully pipelined \verb+rcpps+ instruction
for an approximate reciprocal with reduced accuracy compared to 
a full divide. This approximation is sufficient for this algorithm, and results
in an accuracy similar to GPGPU implementations. An analysis of the impact of the
approximate reciprocal on performance and accuracy is presented in Sect.~\ref{sec:accuracy}.
The
integer cast (line~\ref{l:intc}) is implemented via the vectorized hardware rounding
instruction \verb+roundps+, which was introduced with SSE4.

\begin{figure}[tb]
\centering
\includegraphics*[width=0.9\linewidth]{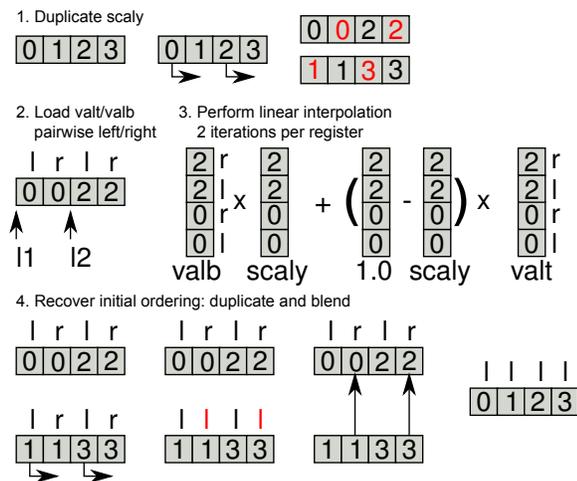}
\caption{Vectorization of part 2 of the algorithm: The data is loaded pairwise into the vector registers. The interpolation  of iterations
0,2 and 1,3  are computed simultaneously. Afterwards the results must be reordered for the second interpolation step.}
\label{fig:vecpart2}
\end{figure}

Part 2 of the algorithm cannot be directly vectorized. Since the pixel coordinates
from step 1 are already in a vector register, the index calculation for, e.g.,
\verb-iv*ISX+iu- and \verb-(iv+1)*ISX+iu- (lines~\ref{l:idx1}, \ref{l:idx2}, 
\ref{l:idx3}, and \ref{l:idx4} in Listing~\ref{lst:alg}) is done using the SIMD floating
point units.  There are pairs of values which can be loaded in one step because
they are consecutive in memory: \verb+valtl+/\verb+valtr+, and
\verb+valbl+/\verb+valbr+, respectively.  
Fig.~\ref{fig:vecpart2} shows the steps involved to
vectorize part 2 and the first linear interpolation. The conversion
of the index into a general purpose register, which is needed for addressing the
load of the data and the scattered pairwise loads, is costly in terms of
necessary instructions. Moreover the runtime
increases linearly with the width of the register, and the whole operation
is limited by instruction throughput.
There are different implementation options with the instructions available
(see below). 
Finally the bilinear interpolation in part 3 is again straightforward to vectorize 
and fully benefits from wider SIMD registers.

We consider two SSE implementations, which only differ in 
part 2 of the algorithm. Version 1 (V1) converts the
floating point values in the vector registers to four quadwords and stores the
result back to memory (cache, actually).  Single index values are then loaded to general purpose
registers one by one.  Version 2 (V2) does not store to memory
but instead shifts all values in turn to the lowest position in the SSE register,
from where they are moved directly to a general purpose register using the 
\verb+cvtss2si+ instruction.
%

Note that any further inner loop unrolling beyond what is required by
SIMD vectorization would not show any benefit due to register shortage;
however, as will be shown later, SMT can be used to achieve a similar 
effect.

\subsection{AVX implementation}

In theory, the new AVX instruction set extension doubles the performance
per core. The backprojection cannot fully benefit from this advantage
because the number of required instructions increases linearly with
the register width in part 2 of the algorithm. For arbitrary SIMD vector
lengths a hardware gather operation would be required to prevent this part from
becoming a severe bottleneck.

Also the limited number of AVX instructions that natively operate on 256-bit 
registers impedes more sophisticated variants of part 2; only the simple
version V1 could be ported. Implementation of V2 would be possible
only at the price of a much larger instruction count, so this alternative
was not considered.
Still an improvement of 25\% could be achieved with the AVX kernel on Sandy Bridge
(see Sect.~\ref{sec:results} for detailed performance results).

Intel already announced the successor AVX2, which will be implemented in the
Intel Haswell processor in 2012.  AVX2 will fix issues preventing better
performance for the backprojection algorithm which are the hardware gather
instruction and a more complete instruction set operating on the full SIMD
register width.


\section{In-depth performance analysis}
\label{sec:model}

\subsection{Analytic performance model}\label{sec:apm}

\begin{figure*}[tb]\centering
    \subfloat[Harpertown]{\includegraphics*[width=0.21\linewidth]{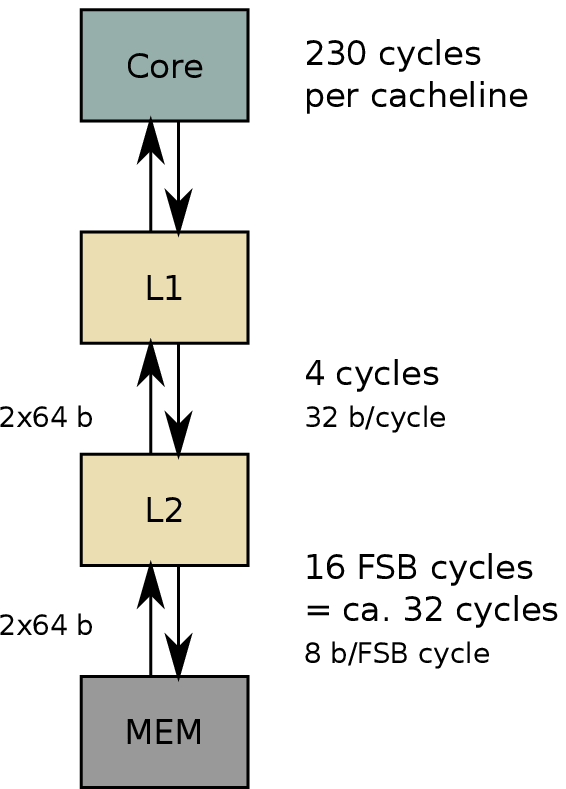}}\hfill
    \subfloat[Westmere]{\includegraphics*[height=0.38\linewidth]{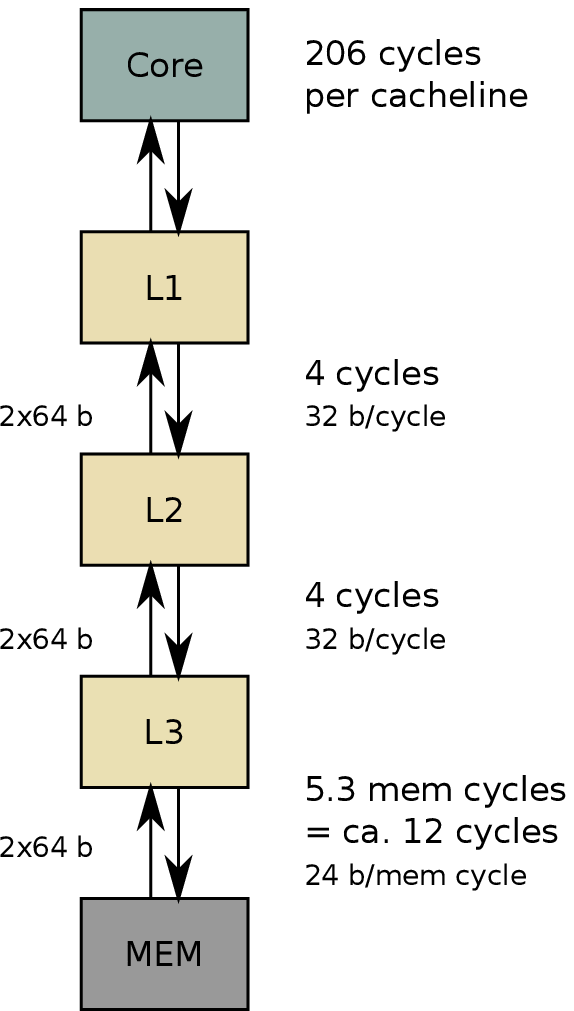}}\hfill
    \subfloat[Sandy Bridge (SSE)]{\includegraphics*[height=0.38\linewidth]{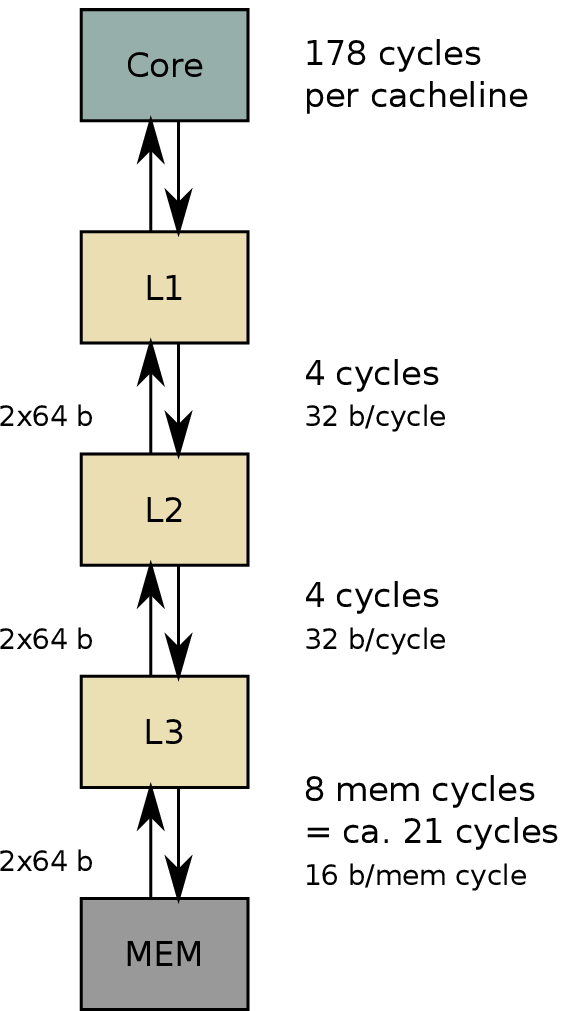}}\hfill
    \subfloat[Westmere EX]{\includegraphics*[height=0.38\linewidth]{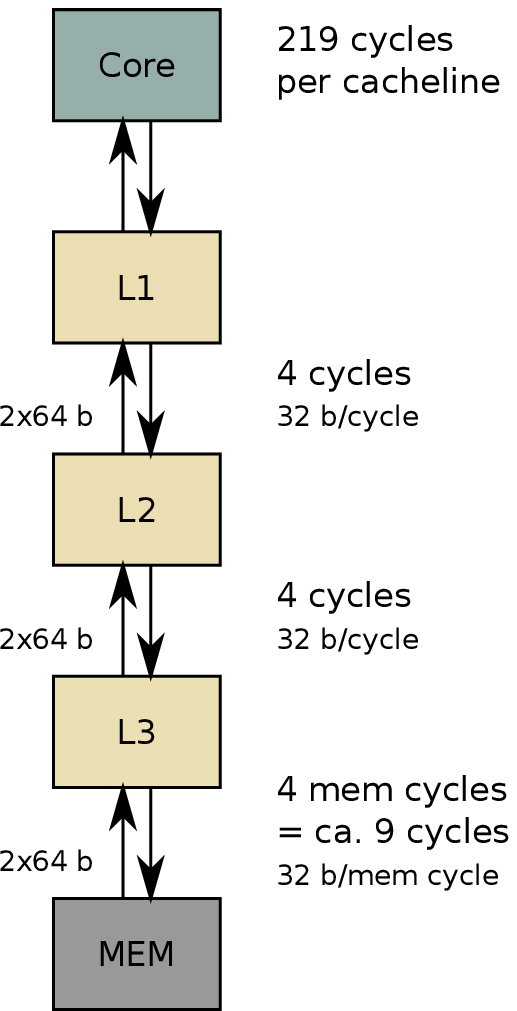}}
    \caption{Performance Analysis: Runtime contributions from instruction execution and necessary cacheline transfers.
    Each arrow is a 64-\BYTE\ cacheline transfer. The total data volume in \BYTES\ is indicated on the left
    of each group of arrows. On the right we show
    the data transfer capabilities between hierarchy levels and the resulting transfer time in core cycles. 
    This assumes that 
    the transfer time is solely determined by bandwidth capabilities, and any latency influence is ignored.
    For data transfers from main memory the contribution in memory/FSB cycles are translated to 
    core cycles. In-core execution times are measured values from Table~\ref{tab:kernel_runtime},
    scaled to a complete cacheline.}
    \label{fig:perf_model}
\end{figure*}
Popular performance models for bandwidth-limited al\-go\-rithms reduce the
influence on the runtime to the algorithmic balance, taking into account the
sustained main memory performance~\cite{hpc4se}. This assumption works well 
in many cases, especially when there is a large gap between arithmetic capabilities and 
main memory bandwidth. Recently it was shown \cite{ppam09} that this simplification
is problematic on newer architectures like Intel Core i7 because of their
exceptionally large memory bandwidth per socket;
e.g., a single Nehalem core cannot saturate the memory interface~\cite{konwihr10}. 
Moreover the additional L3 cache decouples
core instruction execution from main memory transfers and generally provides a better
opportunity to overlap data traffic between different levels of the
memory hierarchy.
Note that in a multithreaded scenario the simple bandwidth model can indeed work
as long as multiple cores are able to saturate the socket
bandwidth. This property depends crucially on the algorithm, of course.

In~\cite{ppam09} we have introduced a more detailed analytic method for cases where 
the balance model is not sufficient to predict performance with the required
accuracy, and the in-cache
data transfers account for a significant fraction of overall runtime. This
model is based on an instruction analysis of the innermost loop body and
runtime contributions of cacheline transfer volumes through the whole
memory hierarchy. We first cover single-threaded execution only and then generalize
to multithreading on the socket once the basic performance limitations
are understood.

A useful starting point for all further analysis is the single-thread
runtime spent
executing instructions with data loaded from L1 cache. The Intel
Architecture Code Analyzer (IACA)~\cite{iaca} was used to analytically
determine the runtime of the loop body. This tool calculates the raw
throughput according to the architectural properties of the processor
under the assumption that all data resides in the L1 cache.
It supports Westmere and Sandy Bridge (including AVX) as target
architectures.
The results for Westmere are shown in the following table for the two
SSE kernel variants described above (all
entries except {\muop}s are in core cycles):
\begin{center}\small
\begin{tabular}{lccccccccc}
          & \multicolumn{6}{c}{Issue port}&   &         &  \\\cline{2-7}  
  Variant &  0  &  1 &  2  &  3 &  4 &  5&  TP& $\mu$OPs&CP\\
  \hline
  V1      & 15  &  21&  \textbf{24} &  3 &  3 & 19&\textbf{24}&85&54\\
  V2      & 20  &  \textbf{27}&  16 &  1 &  1 & 20&\textbf{27}&85&71\\
\end{tabular}
\end{center}
Execution times are calculated separately for all six issue ports
(0\ldots 5). (A \muop\ is a RISC-like ``micro-instruction;'' x86 processors perform
an on-the-fly translation of machine instructions to {\muop}s, which are
the ``real'' instructions that get executed by the core.)
Apart from the raw throughput (TP) and the total number of {\muop}s the
tool also reports a runtime prediction taking into account latencies on the
critical path (CP). Based on this prediction V1 should be faster than V2 on Westmere.
However, the measurements in Table~\ref{tab:kernel_runtime} show
the opposite result. The high pressure on
the load issue port (2) together with an overall high pressure on all ALU issue
ports (0, 1, and 5) seems to be decisive.  In V2 the pressure on port 2
is much lower, although the overall pressure on all issue ports is slightly 
larger.

Below we report the results for the Sandy Bridge architecture with SSE and AVX. 
The pressure on
the ALU ports is similar, but due to the doubled SSE load performance Sandy
Bridge needs only half the cycles for the loads in kernel V1. V1 is therefore
faster than V2 on Sandy Bridge (see Table~\ref{tab:kernel_runtime}).
\begin{center}\small
\begin{tabular}{lccccccccc}
          & \multicolumn{6}{c}{Issue port}&   &         &  \\\cline{2-7}  
  Variant &  0  &  1 &  2   &  3 &  4 &  5& TP& $\mu$OPs&CP\\
  \hline
  V1 SSE  & 16  &  \textbf{20}&  14 &  13 &  3 & 19&\textbf{20}&85&56\\
  V2 SSE  & 20  &  \textbf{26}&  9   &  8  &  1 & 21&\textbf{26}&85&72\\
  V1 AVX  & 18  &  20&  22 &  21 &  6 & \textbf{30}&\textbf{30}&114&90\\
\end{tabular}
\end{center}
\begin{table}[tb]\centering
\begin{tabular}{lcccc}
	              &HPT           &WEM     &WEX            &SNB\\\hline
        V1 SSE    &62.6         &61.6 &59.6        &44.4\\
        V2 SSE    &57.4         &51.5 &54.7        &50.0\\
        V1 AVX    &               &        &               &76.2\\
\end{tabular}
\caption{\label{tab:kernel_runtime}Measured execution times (one core) in cycles 
	for one iteration of the SIMD-vectorized kernel (i.e., 4 or 8 voxel 
	updates) with all operands
  	residing in L1 cache.}
\end{table}

So far we have assumed that all data resides in the L1 cache.  The
data transfers required to bring cachelines into L1 and back to memory
are modeled separately. We assume that there is no overlap between
data transfers and instruction execution. This is true at least for
the L1 cache: It can either communicate with L2 to load or
evict a cacheline, or it can deliver data to the registers, but not
both at the same time. As a first approximation we also
(pessimistically) assume that this ``no-overlapping'' condition holds
for all caches, and that a data transfer between any two adjacent
levels in the memory hierarchy does not overlap with anything else.
Since the smallest transfer unit is a 64-byte cacheline, the analysis
will from now on be based on a full ``cacheline update'' (16 four-byte
voxels), which corresponds to four (two) inner loop iterations when
using SSE (AVX).

We only consider the data traffic for voxel updates; the image data
traffic is negligible in comparison, hence we assume that all image
data comes from L1 cache. It takes two cycles to transfer one cacheline
between adjacent cache levels over the 256-bit unidirectional data path.
Every modified line must eventually be evicted, which
takes another two cycles.
Figure~\ref{fig:perf_model} shows a full analysis, in which the core
execution time for a complete cacheline update is based on the measured cycles
from Table~\ref{tab:kernel_runtime}. 
On the three architectures with L3 cache the simplification is made
that the ``Uncore'' part (L3 cache, memory interface, and QuickPath interconnect) 
runs at the same frequency as the core, which is not strictly true but
does not change the results significantly. It was shown
for the Nehalem-based architectures (Westmere and Westmere EX) that they can overlap
instruction execution with reloading data from memory to the last level 
cache~\cite{konwihr10}. 
Hence, the model predicts that the in-core execution time is much larger than all
other contributions, which makes this algorithm limited by
instruction throughput for single core execution. On Sandy Bridge,
the AVX kernel requires 76.2 cycles for one vectorized loop iteration (eight
updates). This results in 152 cycles instead of 178 cycles (SSE) for one
cacheline update.

Based on the runtime of the loop kernel we can now estimate the total required
memory bandwidth for multithreaded execution if all cores on a socket are utilized, 
and also derive the expected performance (we consider the full volume  without 
clipping):
\begin{center}\small
\begin{tabular}{lccccc}
	                 &HPT        &WEM     &WEX    &SNB  &\multicolumn{1}{m{0.9cm}}{\centering SNB (AVX)}  \\
	    \hline                                               
  BW/core [\GBS]     &1.7          &1.9     &1.5    &2.5  &3.0  \\
  BW/socket [\GBS]   &\textbf{6.8} &11.2    &11.6   &10.0 &12.0 \\
  Perf. [\GUPS]      &0.85         &1.42    &1.45   &1.25 &1.51 \\
\end{tabular}
\end{center}
We conclude that the multithreaded code is bandwidth-limited only on 
Harpertown, since
the required socket bandwidth is above the practical limit given by
the update benchmark (see Table~\ref{tab:arch}). All other
architectures are below their data transfer capabilities for this operation and
should show no benefit from further bandwidth-reducing optimizations 
(see Sect.~\ref{sec:cblock}).

\subsection{ILP optimization and SMT}\label{sec:ilpsmt}

At this point the analysis still neglects the possible benefit from
SMT.  SMT can significantly improve the efficiency of the floating
point units for codes that are limited by
instruction throughput and suffer from underutilization of the
arithmetic units due to dependencies or instruction scheduling issues.
This is definitely the case here, as indicated by the discrepancy
between the ``throughput'' and ``critical path'' predictions in the
previous section. Due to the complex loop body register dependencies
are unavoidable, resulting in many pipeline bubbles. Outer loop
unroll and jam (interleaving two outer loop iterations in the inner
body) is out of the question due to register shortage, but SMT can
do a similar job and provide independent instruction streams using
independent register sets. Since memory bandwidth is no limitation on all
SMT-enabled processors considered here, running two threads on the 
two virtual cores of each physical core is expected to reduce the cycles
taken for the cacheline update.  However, the effect of using SMT is
difficult to estimate quantitatively.
See Sect.~\ref{sec:results} below for complete parallel results



\section{OpenMP parallelization}\label{sec:omp}

OpenMP parallelization of the algorithm is straightforward and works
with all optimizations discussed so far. For the thread counts and
problem sizes under consideration here it is sufficient to parallelize
the loop that iterates over all voxel volume slices 
(loop variable \verb.z. in Listing~\ref{lst:alg}). 
However, due to the clipped-off voxels at the edges and
corners of the volume, simple static loop scheduling with default
chunksize leads to a strong load imbalance. This can be easily
corrected by using block-cyclic scheduling with a small chunksize
(e.g., \verb.static,1.). 

Images
are produced one by one during the C-arm rotation, and could at best 
be delivered to the application in batches. Since the reconstruction
should start as soon as images become available, a parallelization 
across images was not considered.

As shown in Sect.~\ref{sec:model}, the socket-level performance
analysis does not predict strong benefits from bandwidth-reducing
optimizations except on the Harpertown platform. However, since one
can expect to see more bandwidth-starved processor designs with a
more unbalanced ratio of peak performance to memory bandwidth in the
future, we still consider bandwidth optimizations important for this
algorithm. Furthermore, ccNUMA architectures have become omnipresent
even in the commodity market, making locality and bandwidth awareness
mandatory.  In the following sections we will describe a
proper ccNUMA page placement strategy for voxel and image data, and a
blocking optimization for bandwidth reduction. The
reason why we present those optimizations in the context of
shared-memory parallelization is that they become relevant only in the
parallel case, since bandwidth is not a problem on all architectures
for serial execution (see Sect.~\ref{sec:apm}).

\subsection{ccNUMA placement}\label{sec:numa}

The reconstruction algorithm uses essentially two relevant data
structures: the voxel array and the image data arrays. Upon voxel
initialization one can easily employ first-touch initialization, using
the same OpenMP loop schedule (i.e., access pattern) as in the main
program loop. This way each thread has local access (i.e., within its own 
ccNUMA domain) to its assigned voxel layers, and the full aggregate 
bandwidth of a ccNUMA node can be utilized.

Although the access to the projection image data is much less
bandwidth-intensive than the memory traffic incurred by the voxel updates,
ccNUMA page placement was implemented here as well.  As mentioned in
Sect.~\ref{sec:algo_optimization}, the padded projection buffers are explicitly
allocated and initialized in each locality domain, and a local copy is shared
by all threads within a domain. Since the additional overhead for the
duplication is negligible, this ensures conflict-free local access to all image
data. The time taken to copy the images to the local buffers is included in the
runtime measurements.

\subsection{Blocking/unrolling}\label{sec:cblock}

In order to reduce the pressure on the memory interface we use a
simple blocking scheme for the outer loop over all images: 
Projections are loaded and copied to the
padded projection buffers in small chunks, i.e., $b$ images at a time.
The line update kernel (see Sect.~\ref{sec:core_optimization})
for a certain pair of $(y,z)$ coordinates is then
executed $b$ times, once for each projection. This corresponds 
to a $b$-way unrolling of the image loop and a subsequent jam into
the next-to-innermost voxel loop (across the $y$ voxel coordinate).
At the problem sizes
studied here, all the voxel data for this line can be kept in the L1 
cache and reused $b-1$ times. Hence, the complete
volume is only updated in memory $496/b$ instead of 496 times. 
Relatively small unrolling factors between $2$ and $8$ are thus
sufficient to reduce the bandwidth requirements to uncritical
levels even on ``starved'' processors like the Intel Harpertown.
 
This optimization is so effective that it renders proper ccNUMA
placement all but obsolete; we will thus not report the benefit of
ccNUMA placement in our performance results, although it is 
certainly performed in the code.

\section{Results}\label{sec:results}

In order to evaluate the benefit of our optimizations
we have benchmarked different code versions with the $512^3$ case on all test machines.
\rabbitct{} includes a benchmarking application, which takes care of timing and error
checking. It reports total runtime in seconds for the
complete backprojection. We performed additional hardware performance counter
measurements using the likwid-perfctr tool. Likwid-perfctr  can produce 
high-resolution timelines of counter data and useful derived metrics on the core and node
level without changes to the source code. 
Unless stated otherwise we always report results
using two SMT threads per core. For all architectures apart from Sandy Bridge the line update
kernel version V2 was used. On Sandy Bridge results for the SSE kernel V1 as well
as for the AVX port of the V1 kernel are presented.



\subsection{Validation of analytical predictions}
\label{sec:model_validation}

To validate the predicted performance of the analytic  model (see
Sect.~\ref{sec:model}), 
single-socket runs were performed without the 
clipping optimization and SMT. Blocking
was used on the Harpertown
platform only, to ensure that execution is not dominated
by memory access. The following
table shows the measured performance and the deviation against the model
prediction:
\begin{center}\small
\begin{tabular}{m{1cm}ccccc}
                    &HPT  &WEM  &WEX  &SNB  &\multicolumn{1}{m{1cm}}{\centering SNB (AVX)}\\\hline
Perf. [\GUPS]       &0.75 &1.20 &1.30 &1.11 &1.28  \\
deviation           &-13.3\%&-18.3\%  &-11.5\% &-12.6\% &-18.0\%  
\end{tabular}
\end{center}
This demonstrates that the model has a reasonable predictive power. It has been
confirmed that the contribution of data transfers indeed vanishes against the core 
runtime, despite the
fact that the total transfer volume is high and a first rough estimate
based on data transfers and arithmetic throughput alone
(Sect.~\ref{sec:code_analysis}) predicted a bandwidth limitation of this
algorithm on all machines. 

As a general rule, the IACA tool can provide a rough estimate of the innermost
loop kernel runtime via static code analysis. Still it is necessary to further
enhance the machine model to improve the accuracy of the predictions.
Especially the ability of the out of order scheduler to exploit superscalar
execution was overestimated and has led to qualitatively wrong predictions.

Note that this example is an extreme case
with all data transfers vanishing against core runtime. However, the 
approach also works for bandwidth-limited codes, as was shown in~\cite{ppam09}.

\subsection{Impact of optimizations on accuracy}\label{sec:accuracy}

One of the costly operations in this algorithm is the divide involved. A
possible optimization is to use the fully pipelined reciprocal (\verb+rcpps+),
which provides an approximation with 12-bit accuracy compared
to the 24-bit accuracy of the regular divide instruction. The accuracy of the
reciprocal can be improved to at least 21~bits using a Newton-Raphson iteration
\cite{NewtonRaphson}. This requires four additional arithmetic
instructions in the implementation. 
The following table shows the peak signal-to-noise ratio (PSNR) and performance 
of the three alternatives:
Regular divide instruction, reciprocal, and reciprocal with Newton-Raphson iteration, on 
the Westmere test machine for the fully optimized case (full node).
\begin{center}
\begin{tabular}{lcc}
                        &\multicolumn{1}{m{2cm}}{\centering Performance [\GUPS]}    &PSNR [\DB]\\\hline 
         \verb+divps+   &3.79                    &103 \\
         \verb+rcpps+   &4.21                    &67.7  \\
         Newton-Raphson &3.79                    &103 \\
\end{tabular}
\end{center}
As usual in image processing, the PSNR was determined as
the logarithm of a mean quadratic deviation from 
a reference image:
\bq
\mbox{\itshape PSNR} = 10\cdot\log_{10}\frac{M^2}{\displaystyle L^{-3}\sum_{x,y,z}\left[V(x,y,z)-R(x,y,z)\right]^2}
\eq
Here $V(x,y,z)$ is the reconstructed voxel grayscale value at
coordinates $(x,y,z)$ scaled to the interval $[0,M]$, and  
$R(x,y,z)$ is a reference voxel with the ``correct'' result.
The higher the PSNR, the more accurate the reconstruction. 

There is no significant difference neither in performance nor accuracy
between the regular divide and \verb+rcpps+ with Newton-Raphson.  The
performance improvement when using \verb+rcpps+ alone is 10\%, at an
accuracy which is still better than for the published GPGPU
implementations (63--65~\DB). In the following section we discuss
performance results using this latter version.

\subsection{Parallel results}

\begin{figure*}[tb]
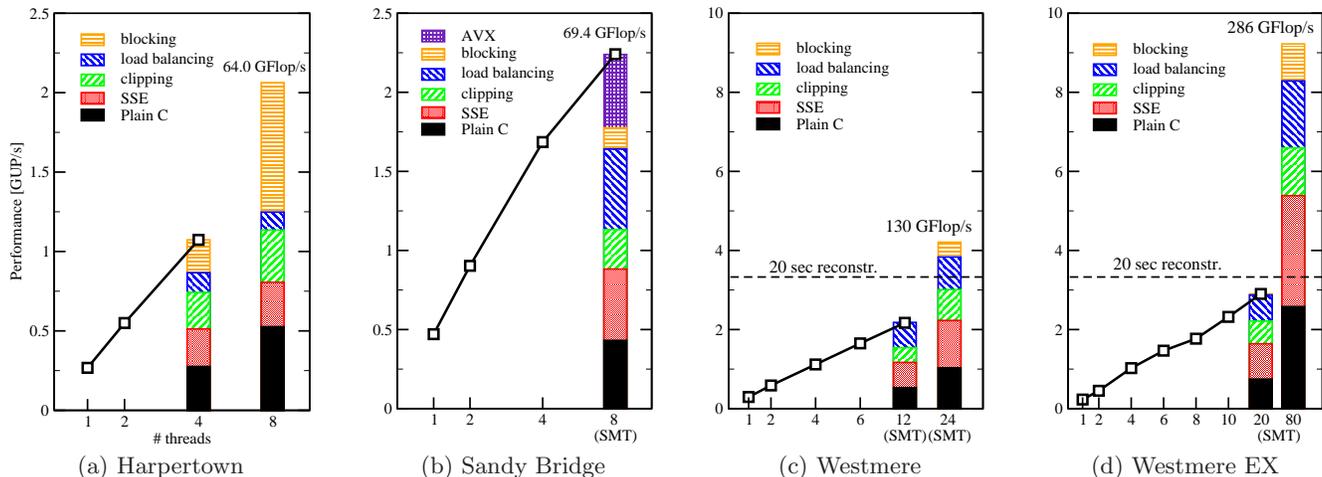

        \subfloat[Harpertown]{
                \includegraphics*[height=0.33\linewidth]{harpertown-perf.eps}
                \label{fig:harpertown-perf}%
        }\hfill
        \subfloat[Sandy Bridge]{
                \includegraphics*[height=0.33\linewidth]{sandy-perf.eps}
                \label{fig:sandy-perf}%
        }\hfill
	\subfloat[Westmere]{%
                \includegraphics*[height=0.33\linewidth]{westmere-perf.eps}%
                \label{fig:westmere-perf}
        }\hfill
        \subfloat[Westmere EX]{
                \includegraphics*[height=0.33\linewidth]{wesex-perf.eps}
                \label{fig:nehex-perf}%
        }
        \caption{Scalability and performance results for the $512^3$ test case
	on all platforms. In-socket scalability was tested
        using the best version of the SIMD-vectorized line update kernel on 
        each system (AVX-V1 on Sandy Bridge, SSE-V2 on all others).
	The practical performance goal for complete reconstruction 
	(20 seconds runtime, corresponding to 3.33\,\GUPS) is indicated 
	as a dashed line. \GFS\ numbers have been computed assuming
        31~{\FLOP}s per optimized (scalar) inner loop iteration.
        Note the scale change between the left and right pairs
	of graphs.\label{fig:allperf}}
\end{figure*}
Figures \ref{fig:allperf}\,(a)--(d) display a summary
of all performance results on node and socket levels, and parallel scaling 
inside one socket for the best version on each architecture. All machines
show nearly ideal scaling inside one socket when using physical cores only. 
With SMT, the benefit is considerable on Sandy Bridge (33\%) and Westmere (31\%),
and a little smaller on Westmere EX (25\%). The large effect on Sandy Bridge
may be attributed to a higher number of bubbles in the pipeline, as indicated by
the larger discrepancy between the ``throughput'' and ``critical path'' cycles
in the AVX loop kernel (see Sect.~\ref{sec:ilpsmt}).
Scalability from one to all sockets of the node is also close to perfect
for the multisocket machines, with the exception of Westmere EX, on which
there is a slight load imbalance due to 80 threads working on only
512 slices. 

\begin{figure}[tbp]
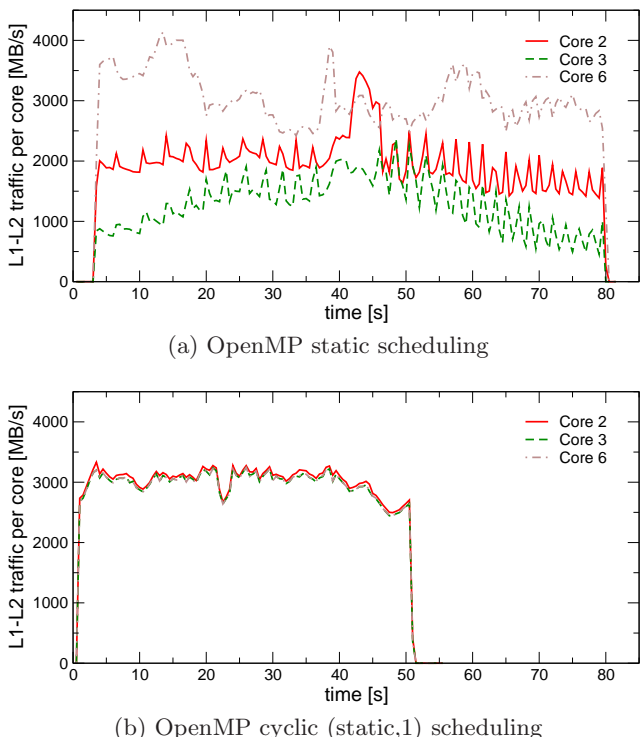

        \subfloat[OpenMP static scheduling]{%
          \includegraphics*[width=\linewidth]{timelineL2static.eps}
                \label{fig:l2s}}\\%
        \subfloat[OpenMP cyclic (static,1) scheduling]{%
          \includegraphics*[width=\linewidth]{timelineL2static1.eps}
                \label{fig:l2s1}}%
\caption{\label{fig:L2timeline}Timeline view of the L2 to L1 cacheline
	traffic for three cores of a 6-thread run without clipping
	on one Westmere socket, using (a) standard static OpenMP
	scheduling and (b) cyclic ``static,1'' scheduling.}
\end{figure}
Depending on the architecture, SSE vectorization boosts performance by
a factor of 2--3 on the socket level. As
explained earlier (see Sect.~\ref{sec:core_optimization}), part 2 of the 
algorithm prohibits the optimal speedup of 4 because its runtime is linear in the
SIMD vector length. Work reduction through clipping alone shows only
limited effect due to load imbalance, but this can be remedied by an appropriate
cyclic OpenMP scheduling, as described in Sect.~\ref{sec:omp}. 
This kind of load balancing not only improves the work
distribution but also leads to a more similar
access pattern to the projection images across all threads.
This can be seen in Fig.~\ref{fig:L2timeline}\,(a), which shows
the cacheline traffic between the L2 and L1 cache during a
6-thread run on one Westmere socket with all optimizations except
clipping (only 3 cores are shown for clarity). Although 
the amount of work, i.e., the number of voxels, is perfectly 
balanced across all threads, there is a strong discrepancy
in cacheline traffic between threads when standard static
scheduling is used. The reason for this is that the projections
of voxel lines onto the detector are quite different for lines
that are far apart from each other in the volume, which leads 
to vastly different access patterns to the image data, and
hence very dissimilar locality properties. Cyclic scheduling
removes this variation (see Fig.~\ref{fig:L2timeline}\,(b)).

\begin{figure}[tb]
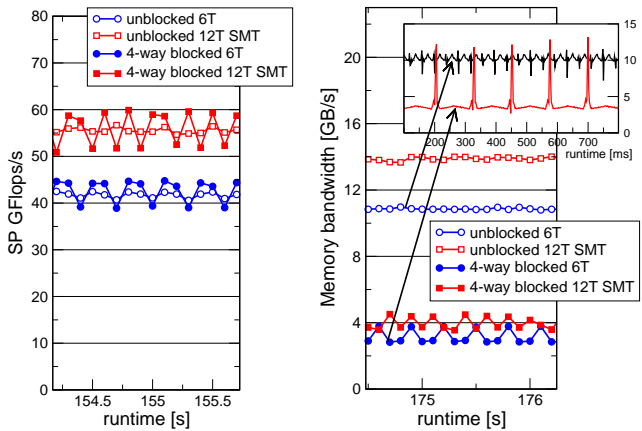
\centering
    \subfloat[FP performance]{\includegraphics*[height=0.67\linewidth]{FLOPS.eps}}\hfill
    \subfloat[Memory bandwidth]{\includegraphics*[height=0.67\linewidth]{MEM.eps}}
    \caption{Performance counter timeline monitoring of floating-point performance (a)
	and  memory bandwidth (b), comparing blocked/nonblocked and SMT/non-SMT
	variants of the best implementation on one Westmere socket at 100\,ms resolution. 
	The inset in (b) shows a zoomed-in view with 2\,ms resolution.}
    \label{fig:perfctr}
\end{figure}

Cache blocking
has little to no effect on all architectures except Harpertown,
as predicted by our analysis. 
Fig.~\ref{fig:perfctr} shows timelines for socket floating point
performance and bandwidth on one Westmere socket, comparing
blocked/nonblocked and SMT/non-SMT variants.
Fig.~\ref{fig:perfctr}\,(a) clearly demonstrates the performance boost
of SMT in contrast to the very limited effect of blocking.
On the other hand, the blocked implementation significantly reduces the
bandwidth requirements (Fig.~\ref{fig:perfctr}\,(b)). 
The blocked variants have a noticeable amplitude of variations while the 
nonblocked versions appear
smooth. In the inset of Fig.~\ref{fig:perfctr}\,(b) we show a zoomed-in
view with finer time resolution, which indicates that the frequency
of bandwidth variations is much larger without blocking; this is 
plausible since the bandwidth is dominated by the voxel volume
in this case. With blocking, individual voxel lines are transferred
in ``bursts'' with phases of inactivity in between, where image data
is read at low bandwidth.

The benefit of AVX on Sandy Bridge falls short of expectations
for the same reason as in the SSE case. Still it is remarkable 
that the desktop Sandy Bridge system
outperforms the dual-socket Harpertown server node, which features twice
the number of cores at a similar clock speed. Both Westmere and
Westmere EX meet the performance requirements of at most 20\,sec for a complete
volume reconstruction. The Westmere EX node is, however, not competitive 
due to its unfavorable price to performance ratio. It is an option if
absolute performance is the only criterion. 

\section{CPU vs. GPGPU}\label{sec:gpgpu}

\begin{figure}[tb]\centering
\includegraphics*[width=0.99\linewidth]{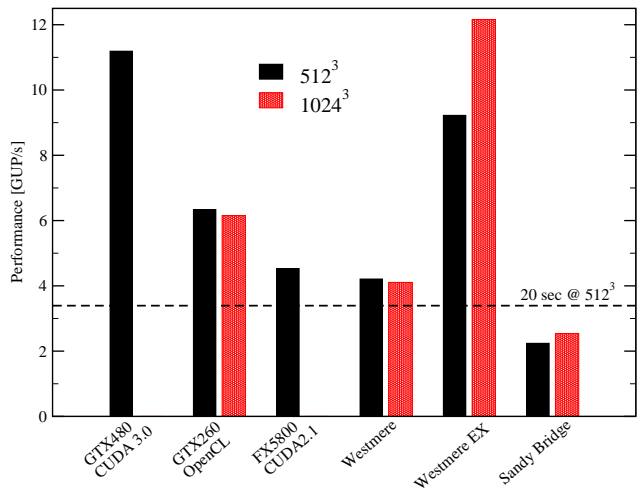}
\caption{\label{fig:gpgpu}Performance comparison between the best
	reported GPGPU implementations in OpenCL and CUDA and our
	CPU versions on the systems in the test bed for problem sizes
	$512^3$ and $1024^3$, respectively. There is currently no working
	CUDA implementation for the latter case. The practical 
	performance goal for complete reconstruction of a $512^3$ volume
	(3.33\,\GUPS) is indicated by a dashed line.}
\end{figure}
Since the backprojection algorithm is well suited for GP\-GPUs, the
performance leaders of the open competition benchmark have traditionally been 
GPGPU implementations in OpenCL and CUDA. 
The gap to the fastest CPU version reported on the \rabbitct{}
Web site~\cite{hpc:RabbitCT:2011} before this work was started
was very large. For the reasons given in the derivation of the performance
model, simple bandwidth or peak performance comparisons are
inadequate to estimate the expected reconstruction speed advantage
of GPGPUs, although it should certainly lie inside the usual corridor
of 4--10 when comparing with a full CPU socket. 
An example of a well-optimized GPU code was published 
recently~\cite{papenhausen11}.
Our implementation shows that current x86 multicore chips are
truly competitive (see Fig.~\ref{fig:gpgpu}), 
even when considering the price/performance ratio.  Beyond
the clinically relevant $512^3$ case, industrial applications need higher
resolutions. This is a problem for GPGPU implementations because the local
memory on the card is often too small to hold the complete voxel volume,
causing extra overhead for moving partial volumes in and out of the 
local memory and leading to a more complex implementation.
If the price for the
hardware is unimportant, a Westmere EX node is an option that 
can easily outperform GPGPUs. Note that the unusually large gap between 
the performance levels at $512^3$ and $1024^3$ on this architecture
may be attributed to better load balancing when 80 threads work on
1024 instead of 512 slices. It also lifts the performance
efficiency (fraction of node peak) on this system to about 
50\%, matching the level achieved by the other multicore 
nodes already on the smaller problem. 
This compares to 26\% for the CUDA code on the GTX480
card, assuming the same number of \FLOP{}s per voxel update. 


The results for the Sandy Bridge desktop system and the good scalability
even of the optimized algorithm promise even better performance levels
on commodity multicore systems in the future.
Note that it would be possible to provide a simple and efficient
distributed memory parallelization of the algorithm for even faster
reconstruction.  ``Micro-clusters'' based on cheap desktop technology
could then easily meet any time constraint at extremely low cost.
However, a comparison based on the hardware cost alone is certainly
too simplistic, especially when taking into account the overall cost
for a complete CT scanner including maintenance.


\section{Conclusions and Outlook}

We have demonstrated several algorithmic and low-level optimizations for a CT
backprojection algorithm on current Intel x86 multicore processors. 
Highly optimizing compilers were not able to deliver
useful SIMD-vectorized code.
Our implementation is thus based on assembly language and vectorized using the
standard instruction set extensions SSE and AVX.
The results
show that commodity hardware can be competitive with highly tuned GPU
implementations for the clinically relevant $512^3$ voxel case at the same level
of accuracy. Nonpipelined divide instructions (\verb.divps.) or a fast 
pipelined version (\verb.rcpps.) with subsequent Newton-Raphson iteration provide 
better accuracy at a 10\% performance penalty.   Our
results showed that it is necessary to consider all aspects of processor and
system architecture in order to reach best performance, and that the effects of
different optimizations are closely connected to each other.  The benefit of
the AVX instruction set on Sandy Bridge was limited due to the lack of a
gathered load and the small number of instructions that natively operate on the
full SIMD register width.  This relevant algorithm can achieve very good
efficiencies on commodity processors and it would  be a natural step to further
improve performance with a distributed memory implementation.
At higher
resolutions, which are used in industrial applications, multicore systems are
frequently the only choice (apart from expensive custom solutions).

Future work includes a more thorough analysis and optimization of the
AVX line update kernel, and an inclusion of the new AVX2 gather
operations once they become available. 

\section{Acknowledgments}
We are indebted to Intel Germany for providing test systems and early access
hardware for benchmarking. This work was supported by the Competence
Network for Scientific High Performance Computing in Bavaria (KONWIHR)
under project OMI4papps.

\bibliographystyle{drgh}
\bibliography{sigproc}
\end{document}